\documentclass[prl,twocolumn,superscriptaddress,showpacs,preprintnumbers,amsmath,amssymb]{revtex4}
\usepackage{graphicx}% Include figure files
\usepackage{dcolumn}% Align table columns on decimal point
\usepackage{bm}% bold math

\def\altwo{Al{\sc \,ii}}

\def\approxgt{\mathrel{\spose{\lower 3pt\hbox{$\sim$}}
        \raise 2.0pt\hbox{$>$}}}
\def\approxlt{\mathrel{\spose{\lower 3pt\hbox{$\sim$}}
        \raise 2.0pt\hbox{$<$}}}

\def\catwo{Ca{\sc \,ii}}

\def\cone{C{\sc \,i}}

\def\crtwo{Cr{\sc \,ii}}
\def\cs{$\chi^2$}
\def\csn{$\chi^2/\nu$}
\def\daa{$\Delta \alpha/\alpha$}
\def\dxx{$\Delta x/x$}

\def\dxxavq{$\langle\Delta x/x\rangle_{\rm abs}$}

\def\dxxavtw{$\langle\Delta x/x\rangle^{\rm weighted}_{\rm total}$}

\def\fetwo{Fe{\sc \,ii}}
\def\fr{Fig.~\ref}
\def\kmps{km s$^{-1}$}

\def\mgone{Mg{\sc \,i}}
\def\mgtwo{Mg{\sc \,ii}}
\def\mntwo{Mn{\sc \,ii}}

\def\nitwo{Ni{\sc \,ii}}

\def\sitwo{Si{\sc \,ii}}	
\def\spose#1{\hbox to 0pt{#1\hss}} 
\def\stwo{S{\sc \,ii}}

\def\titwo{Ti{\sc \,ii}}

\def\tr{Table~\ref}

\def\zav{$\langle z\rangle$}

\def\zem{$z_{\rm em}$}
\def\zntwo{Zn{\sc \,ii}}

\def\zuv{$z_{\rm UV}$}
\def\zrad{$z_{\rm 21}$}

\def\znf{Q0952$+$179}
\def\oot{Q1127$-$145}
\def\ott{Q1229$-$021}
\def\ztt{Q0235$+$164}
\def\zet{Q0827$+$243}
\def\otto{Q1331$+$170}
\def\oof{Q1157$+$014}
\def\zff{Q0458$-$0203}

\def\aap{A\&A}

\def\aj{Astron.~J.}

\def\apj{Astrophys.~J.}

\def\aplett{Astrop.~Lett.}

\def\mnras{Mon.~Not.~R.~Astron.~Soc.}
\def\nat{Nature}

\def\prd{Phys.~Rev.~D}

\def\prl{Phys.~Rev.~Lett.}

\def\rmp{Rev.~Mod.~Phys.}

\begin{document}

\preprint{APS/000-AAA}

\title{Limits on variations in fundamental constants
from 21-cm and ultraviolet quasar absorption lines}% Force line breaks with \\

\author{P.~Tzanavaris}
% \altaffiliation[Also at ]{Physics Department, XYZ University.}%Lines break automatically or can be forced with \\
 \email{pana@phys.unsw.edu.au}
\affiliation{%
School of Physics, The University of New South Wales,\\
Sydney, NSW 2052, Australia
}%
\author{J.~K.~Webb}%
% \email{jkw@phys.unsw.edu.au}
\affiliation{%
School of Physics, The University of New South Wales,\\
Sydney, NSW 2052, Australia
}%
\author{M.~T.~Murphy}
% \email{flambaum@phys.unsw.edu.au}
\affiliation{%
Institute of Astronomy, Madingley Road, Cambridge CB3 0HA, United Kingdom
}%
\author{V.~V.~Flambaum}
\affiliation{%
School of Physics, The University of New South Wales,\\
Sydney, NSW 2052, Australia
}%
\author{S.~J.~Curran}
\affiliation{%
School of Physics, The University of New South Wales,\\
Sydney, NSW 2052, Australia
}%

\date{\today}% It is always \today, today
             %  but any date may be explicitly specified

\begin{abstract}
Quasar absorption spectra at 21-cm and UV rest-wavelengths
are used to estimate the time
variation of $x\equiv \alpha^2 g_p \mu$, where $\alpha$ is the fine
structure constant, $g_p$ the proton $g$ factor, and
$m_e/m_p\equiv\mu$ the electron/proton mass ratio.
Over a redshift range $0.24\approxlt z_{\rm abs}\approxlt 2.04$,
\dxxavtw~$=(1.17\pm1.01)\times10^{-5}$.
A linear fit gives
$\dot{x}/x=(-1.43\pm1.27)\times 10^{-15} {\rm yr}^{-1}$.
Two previous results on varying $\alpha$ yield the strong limits
$\Delta\mu/\mu=(2.31\pm1.03)\times 10^{-5}$ and
$\Delta\mu/\mu=(1.29\pm1.01)\times 10^{-5}$.
Our sample, 8$\times$ larger than any previous,
provides the first direct estimate of the intrinsic
21-cm and UV velocity differences $\sim 6$~\kmps.
%
%21-cm and UV quasar absorption spectra are used to estimate the time
%variation of $x\equiv \alpha^2 g_p \mu$, where $\alpha$ is the fine
%structure constant, $g_p$ the proton $g$ factor, and
%$m_e/m_p\equiv\mu$ the electron/proton mass ratio. Our sample is 8
%times larger than previously used. 
%The intrinsic line-of-sight velocity difference
%between 21-cm and UV absorption redshifts, $z_{\rm abs}$, is estimated
%to be $\sim 6$~\kmps. For $0.24\approxlt z_{\rm abs}\approxlt 2.04$,
%\dxxavtw~$=(1.17\pm1.01)\times10^{-5}$. A linear fit gives
%$\dot{x}/x=(-1.43\pm1.27)\times 10^{-15} {\rm yr}^{-1}$.  Using two
%previous results on varying $\alpha$ yields the best limits
%$\Delta\mu/\mu=(2.31\pm1.03)\times 10^{-5}$ and
%$\Delta\mu/\mu=(1.29\pm1.01)\times 10^{-5}$.
\end{abstract}

% PACS, the Physics and Astronomy
                             % Classification Scheme.
%\keywords{Suggested keywords}%Use showkeys class option if keyword
                              %display desired
\maketitle

%%%%%%%%%%%%%%%%%%%%%%%%%%%%%%%%%%%%%%%%%%%%%%%%%%%%%%%%%%%%%%%%%%%%%
The existence of extra spatial dimensions, often invoked by
superunification theories, may be inferred by the detection of spatial
or temporal variations in the values of coupling constants
(See \cite{2003UzanRMP} for a review).
Spectroscopy of gas clouds which intersect the lines of sight to
distant quasars is a unique tool, probing the values of
these constants over a large fraction of the age of the universe. 
The highly sensitive many-multiplet method, developed by
\cite{1999PhRvL..82..888D,1999PhRvL..82..884W}, has been
applied to rest-frame ultraviolet (UV) atomic
quasar absorption lines
to provide constraints on the possible variation of the fine
structure constant, $\alpha\equiv e^2/(\hbar
c)$ \cite{1999PhRvL..82..884W,2001PhRvL..87i1301W,2003MNRAS.345..609M,2004A&A...417..853C,2004PhRvL..92l1302S}.
Molecular hydrogen absorption lines have provided constraints
on the variability of the electron-to-proton mass ratio, $\mu$
\cite{1975ApL....16....3T,1988ApJ...324..267F,1993JETPL.....58..237V,2002AstL...28..423I,2004PRL...92...101302U}.

\paragraph*{Principle.---}The above results have been obtained by
use either of heavy element transitions, which absorb in the
rest-frame UV, or of molecular hydrogen transitions. Another approach
is to use the parameter $x\equiv \alpha^2 g_p m_e/m_p$ when, as well
as rest-frame UV, rest-frame 21-cm absorption, due to cold neutral
hydrogen, is also detected. Rest-frame UV
absorption is observed redshifted in the optical region, and
rest-frame 21-cm is observed redshifted at longer radio wavelengths.
%????
The ratio of frequencies $\omega_{21}/\omega_{\rm UV}\propto x$.
21-cm absorption occurs in a few damped Lyman--$\alpha$
(DLA) systems which also show heavy-element absorption in the UV. A
detailed list can be found in \cite{2005MNRAS.356.1509C}.
If both UV and 21-cm
absorption occur at the same physical location,
the relative change of the value of $x$ between redshifts $z$ and 0 is
related to the observed absorption redshifts for rest-frame
21-cm and UV, \zrad\ and \zuv, according to \dxx~$\equiv
\frac{x_z-x_0}{x_0} = \frac{z_{\rm UV}-z_{\rm 21} } {1+z_{\rm 21}}$
%=\frac{v_{\rm UV} - v_{21}}{c}
\cite{1980ApJ...236L.105T}.  
We obtained values for \zuv\ and \zrad\ by using the strongest
absorption components in an absorption system. This approach is discussed in detail later.
%This was done to increase the probability that
%21-cm and UV absorption are spatially coincident.

However, as there are only 17 DLAs where both 21-cm and UV absorption
have been detected, there are few results based on this method
\cite{1979AJ.....84..699W,1980ApJ...236L.105T,1995ApJ...453..596C,1981ApJ...248..460W}.
Out of these, only \cite{1995ApJ...453..596C} use high-resolution
optical data from the Keck telescope's HIRES spectrograph, but they
provide an estimate of \dxx\ at a single redshift from a single
absorption system.  We applied this method to eight absorption systems
in eight quasar spectra (one system per spectrum), covering the
absorption redshift range $\sim 0.24$ to $\sim 2.04$. We used all
available 21-cm absorption data in conjunction with the
highest-resolution UV data available. Thus the results presented here
are based on the largest dataset of the highest quality to which this
method has been applied to date.

\paragraph*{Data analysis.---}
Details of the 21-cm and UV data used are given in \tr{tab:data}.  All
redshifts are in the heliocentric frame.  For the strongest component
in each 21-cm absorption complex, the dispersion coordinate at the
pixel of minimum intensity, MHz or \kmps, was measured, from which
\zrad\ was obtained.  We searched the optical data for heavy element
absorption features close to the redshifts where there is 21-cm
absorption.  A number of UV absorption features were thus identified,
some due to neutral species and most due to singly ionized species.
For all UV spectra we determined the value of the dispersion
coordinate, \AA\ or \kmps, for the strongest component at the pixel of
minimum intensity.
We then determined absorption redshifts
for each detected neutral or singly ionized absorption species that
was not saturated. A \zuv\ value was determined individually for each
transition of a single species, e.g. independently for
\zntwo~2026.14 and \zntwo~2062.66.  In all, there were 30 distinct UV
species identifications (see \tr{tab:data}, column 6). Detailed
velocity plots showing all 21-cm and UV absorption components used can
be found in \cite{2004Tzan} and \cite{2005Tzan}.

%In the single case of \oot, there was no clear indication which of the
%UV components with \zuv\ closest to \zrad\ was the strongest.  We detected three
%\mntwo\ transitions, two \catwo\ transitions and one \mgone\
%transition.  The strongest component in each of the three \mntwo\
%multiplet transitions was within $\sim 4$~\kmps\ of the strongest 21-cm
%component.  However, the strongest component in each of the two
%\catwo\ multiplet transitions was $\sim 20$~\kmps\ away from the
%strongest 21-cm component.  In the case of \mgone\ both components
%appeared to be of equal strength.  There was no such ambiguity in any
%of the other spectra in our sample.  For this reason, we carried out
%the \zuv--\dxx\ determination twice in this case. The first time we
%assumed that the component corresponding to \zrad\ was the one which
%appeared strongest in \mntwo. The second time we assumed this for the
%component which appeared strongest in \catwo. We then took the average
%of the two results for redshift and \dxx. We note that, in any case,
%this does not significantly affect our final result.

%%%%%%%%%%%%%%%%%%%%%% TABLE %%%%%%%%%%%%%%%%%%%%%%%%%%%%%%%%%%%%%%%%%%%%%%%%%%%%%%%%%%%%%%%%%%%%%%
\begin{table*}
\caption{\label{tab:data}Data used in this work. There is one 21-cm/UV
absorption system in each quasar spectrum. Column 1 is the quasar name
and Column 2 its emission redshift.  Column 3 gives the 21-cm
absorption redshift (and error from Column 4 references) for the
strongest component. We determined this after digitizing 21-cm
absorption plots (references in column 4). For \zff\ the original data
were used with the error taken from \cite{1985ApJ...294L..67W}. Column
5 gives the mean absorption redshift (and standard deviation on the
observed mean) for the strongest UV component.  Column 6 gives the UV
heavy element species observed in the optical (with number of
transitions in parentheses, if more than one). Column 7 gives the
source for the UV data. Data for eight quasars were obtained from the
European Southern Observatory's (ESO) archive and were originally
observed with the UVES spectrograph on the Very Large Telescope (VLT),
in which case the ESO program ID is given in column 7 and the
principal investigators of the program are given in footnotes.  For
\otto\ we also used \sitwo\ 1808.01\AA\ Keck/HIRES data provided by
A.~Wolfe.  For quasar \ztt\ we digitized an absorption plot from the
literature for a single heavy element species.}
\begin{ruledtabular}
\begin{tabular}{cclccll}
Quasar & \zem\ & \zrad\ & 21-cm data        &$\langle z_{\rm UV}\rangle$& ions & UV data \\
\hline
\znf&   1.478 & 0.237803(20)  &\cite{2001AA...369...42K} & 0.237818(6) &\mgone, \catwo(2)                       & 69.A-0371(A)\footnotemark[1] \\
\oot&   1.187 & 0.312656(50)  &\cite{2001MNRAS.325..631K}& 0.312648(6) &\catwo(2), \mntwo(3)            & 67.A-0567(A)\footnotemark[2], 69.A-0371(A)\footnotemark[1]\\
\ott&   1.038 & 0.394971(4)   &\cite{1979ApJ...230L...1B}& 0.395019(40) &\catwo(2), \mntwo(3), \titwo            & 68.A-0170(A)\footnotemark[3]  \\
\ztt&   0.940 & 0.523874(100) &\cite{1978ApJ...222..752W}& 0.523829(6) &\mgone                                  & \cite{1992ApJ...391...48L} \\
\zet&   0.941 & 0.524757(50)  &\cite{2001AA...369...42K} & 0.524761(6) &\catwo(2), \fetwo\                      & 68.A-0170(A)\footnotemark[3], 69.A-0371(A)\footnotemark[1]\\
\otto&  2.097 & 1.776427(20)  &\cite{wolfep}             & 1.776355(5) &\mgone, \altwo, \sitwo, \stwo,          & 67.A-0022(A)\footnotemark[4], 68.A-0170(A)\footnotemark[3]\\
    &         &           &                          &          &\cone(3), \cone$^{\ast}$, \crtwo(2), \mntwo(2),& \\
    &         &           &                          &          &\fetwo(4), \nitwo(6), \zntwo\                      & \\
    &         &           &                          &          &\sitwo\footnotemark[5]                  &\cite{wolfep} \\ 
\oof&   1.986 & 1.943641(10)  &\cite{1981ApJ...248..460W}& 1.943738(3) &\mgone, \mgtwo(2), \sitwo,              & 65.O-0063(B)\footnotemark[6], 67.A-0078(A)\footnotemark[6], \\
    &         &           &                          &          & \nitwo(6)                              & 68.A-0461(A)\footnotemark[7]\\
\zff&   2.286 & 2.039395(80)  &\cite{wolfep}         & 2.039553(4) &\zntwo(2), \nitwo(6), \mntwo(3),        & 66.A-0624(A)\footnotemark[6], 68.A-0600(A)\footnotemark[6], \\
    &         &           &                          &          &\crtwo(3)                               & 072.A-0346(A)\footnotemark[6], 074.B-0358(A)\footnotemark[8] \\
\end{tabular}
\end{ruledtabular}
\footnotetext[1]{Savaglio}
\footnotetext[2]{Lane}
\footnotetext[3]{Mall{\'e}n-Ornelas}
\footnotetext[4]{D'Odorico}
\footnotetext[5]{Same transition as in the UVES data but from Keck/HIRES.}
\footnotetext[6]{Ledoux}
\footnotetext[7]{Kanekar}
\footnotetext[8]{Dessauges-Zavadsky}
\end{table*}

%%%%%%%%%%%%%%%%%%%%%%%%%%%%%%%%%%%%%%%%%%%%%%%%%%%%%%%%%%%%%%%%%%%%%%%%%%%%%%%%%%%%%%%%%%%%%%%%%%

\paragraph*{Estimating \dxx.---}
For each absorption system we calculated $\langle z_{\rm UV}\rangle$,
the average of all UV absorption redshifts for single
species (column 5 in \tr{tab:data}).
Using this and our measured \zrad\ (column 3 in \tr{tab:data})
we applied the relation between \dxx, \zuv\ and \zrad\ to obtain 
\dxxavq. This is the average value for \dxx\ for each absorption
system. 
We plot these results in \fr{fig:dxxz}.

Taking the average of all \dxxavq\ values,
we obtained 
$\langle\Delta x/x\rangle_{\rm total} = (0.91\pm1.04)\times10^{-5}$
(\lq result 1\rq) over an absorption redshift range
$0.24\approxlt z_{\rm abs}\approxlt 2.04$
and a fractional lookback time range
$0.20\approxlt t_{\rm flb} \approxlt 0.76$, where we
have used  a Hubble parameter $H_0 = 73$~\kmps\ Mpc$^{-1}$, a total matter density
$\Omega_M=0.27$ and a cosmological constant 
$\Omega_{\Lambda}=0.73$. 
The error quoted is the standard deviation on the mean.

%However, the value of \dxxavq\ for \zff\ is $5.2\sigma$ away from
%the mean obtained for all objects. Such a large deviation cannot be a
%statistical fluctuation but probably means that 21-cm absorption
%occurs in a velocity component which is clearly different from the UV
%absorption component.  Therefore, we exclude this object from the
%sample and perform the above calculations again.  This leads to the
%quoted result \dxxavt~$=(-0.49\pm0.84)\times10^{-5}$ over the redshift
%range $0.24\approxlt z_{\rm abs}\approxlt 1.94$.  From these values
%and the average lookback time, $\langle t_{\rm lb}\rangle = 6.25 \
%{\rm Gyr}$, we calculate $ \langle\Delta x/x\rangle_{\rm total} /
%\langle -t_{\rm lb}\rangle_{\rm total} =(-0.57\pm1.74)\times 10^{-15}
%{\rm yr}^{-1}$.

We also performed the above calculations by taking into account 
statistical errors on \zrad\ and $\langle z_{\rm UV}\rangle$ per
absorption system (\tr{tab:data}). For \zrad\ we used errors
from the \tr{tab:data} references
\footnote{There are no errors for \oot\ (we used the error from
\cite{2001AA...369...42K} as the same instrument was used by the same
observers).  }. For $\langle z_{\rm UV}\rangle$ we used the standard
deviation on $\langle z_{\rm UV}\rangle$ for each absorption system
\footnote{For the single transition in \ztt, \znf\ and \zet\ we used $6\times 10^{-6}$.
This is the maximum error of all other $\langle z_{\rm UV}\rangle$ except for
the atypical value for \ott.}. We obtained \dxxavtw~$=
(2.18\pm0.97) \times 10^{-5}$ (\lq result 2\rq).  This has a \cs\ per
degree of freedom, $\nu$, \csn~$=8$. We thus increased the individual
errors on \dxxavq\ by an additional error, $s$, to
$(\sigma_{\langle\Delta x/x\rangle_{\rm abs}}^2 + s^2 )^{0.5}$
until, at $s=1.90\times 10^{-5}$, \csn~$=1$. This corresponds to
\dxxavtw~$= (1.17\pm1.01) \times 10^{-5}$ (\lq result 3\rq).

For all $\langle z_{\rm UV}\rangle$ values per quasar absorber we
calculated the average fractional
lookback time per absorber, $\langle t_{\rm flb}\rangle_{\rm abs}$. We
then performed an iterative linear least squares fit to
\dxxavq~$=A\langle t_{\rm flb}\rangle_{\rm abs}$, where the additional
error, $s$, was determined at each iteration to force \csn~$=1$ around
the fit, obtaining $A=(1.90\pm1.69)\times 10^{-5}$\footnote{We adopt
$\Delta x/x =0$ at $z=0$ as the terrestrial value. This has not been
checked elsewhere within the Galaxy, and should thus be taken as an
assumption.}. It follows that the best fit rate of change of $x$ as a
function of time $\frac{d}{dt}(\Delta x/x) = \dot{x}/x_0 =
(-1.43\pm1.27)\times 10^{-15} {\rm yr}^{-1}$.

%%%%%%%%%%%%%%%%%%%%%%%%%%%%%%%% FIGURE 1 %%%%%%%%%%%%%%%%%%%%%%%%%%%%%%%%%%%%%%%%%%%%%%%%%%%%%%%%%%%%%%%%%%%%%%%%%%%%%
\begin{figure}[htb]
\hspace{-1cm}
\includegraphics[width=1\columnwidth, angle=-90]{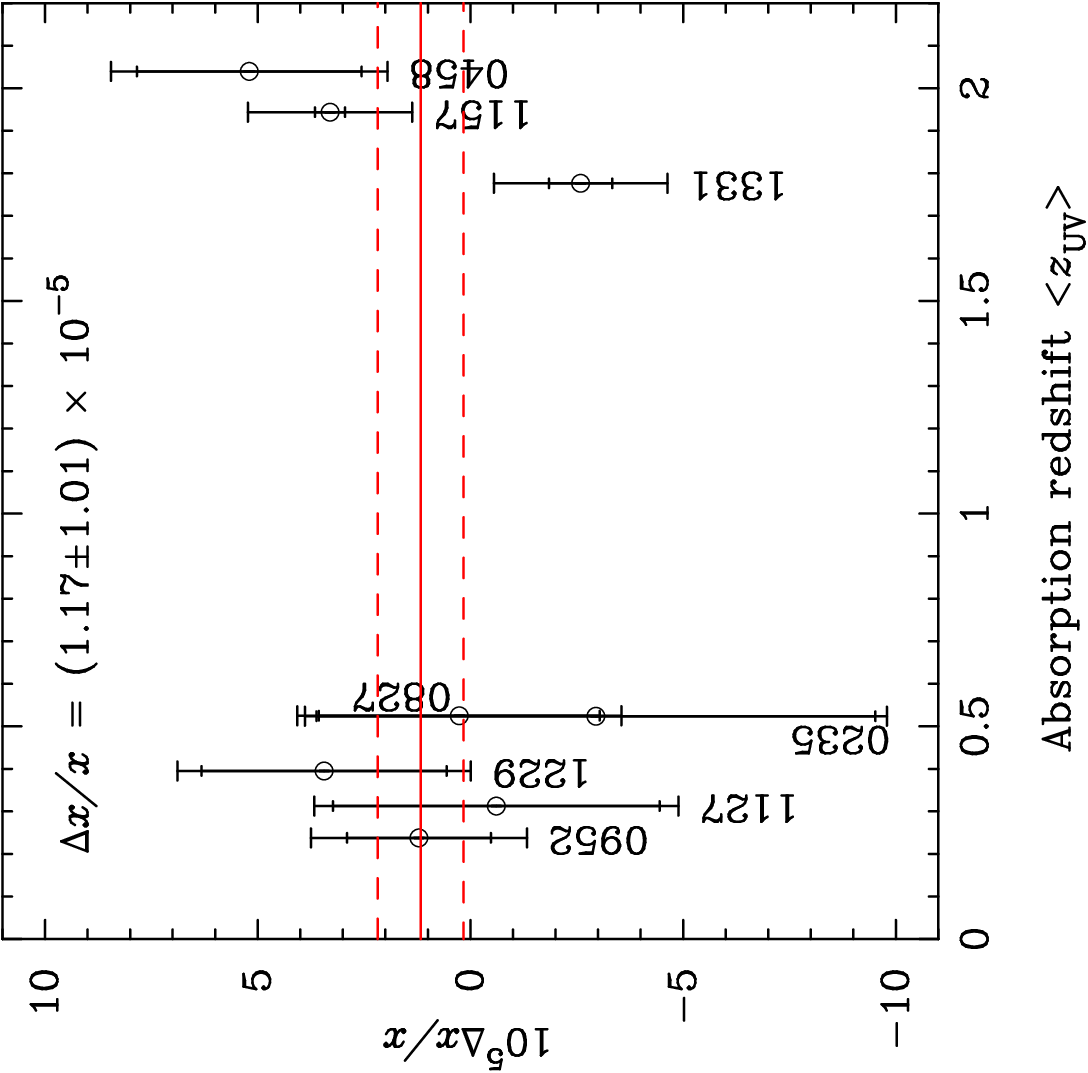}
\caption{\label{fig:dxxz} \dxx\ results for the eight abosrption
systems in our quasar sample. Initial (increased) error bars have
shorter (longer) terminators.  Quasar names are given truncated to
four digits. Each point represents \dxxavq\ obtained from \zrad\ and
$\langle z_{\rm UV}\rangle$, for all heavy element species in a quasar
spectrum, versus average $\langle z_{\rm UV}\rangle$ for that spectrum.
The solid horizontal line is result 3. The dashed lines show the $\pm
1\sigma$ range.}
\end{figure}

%%%%%%%%%%%%%%%%%%%%%%%%%%%%%%%%%%%%%%%%%%%%%%%%%%%%%%%%%%%%%%%%%%%%%%%%%%%%%%%%%%%%%%%%%%%%%%%%%%%%%%%%%%%%%%%%%%%%%%%

\paragraph*{Assumptions.--}
In this work we are making two assumptions.  The validity of our
result does not depend on the validity of these assumptions. On the
contrary, we are essentially using our result to test these
assumptions. Moreover, if these assumptions are incorrect,
they only contribute to any observed scatter in \dxxavq.
Therefore, this possibility has already been taken into account
in results 1 and 3.

1.{\it Strongest components:} Both the 21-cm and UV profiles exhibit 
complex velocity structure,
i.e. have multiple absorption components at slightly different
redshifts.  That being the case, how does one compare the redshifts
among different transitions?  For neutral and singly ionized UV
species, velocity structure is essentially the same and corresponding
components can easily be identified. This is not the case if one
compares 21-cm and UV velocity structure, although we have not
systematically searched for corresponding velocity patterns in the
21-cm and UV profiles. One simple option is to measure mean absorption
centroids over all components for each absorption system, which was essentially the
approach taken by \cite{1995ApJ...453..596C}.  Alternatively one can
simply determine \zrad\ for the strongest component in the 21-cm
profile and \zuv\ for the strongest component in each of the UV
transitions, and use these.  In our sample this is the only well
defined quantity for both the 21-cm and the UV profiles
%\footnote{The only exception to this is \oot: There is good correspondence
%in velocity space between strongest components in all
%UV transitions except \catwo.}.
Additionally, this procedure, unlike mean centroiding, should have the advantage
of being less sensitive to measurement errors caused by
broad velocity structure in absorption lines.
%Additionally,
%this procedure should have the advantage of reducing measurement
%errors caused by broad velocity structure in absorption lines.
%????
%????

2.{\it Neutral and singly ionized species:}
We have used both neutral and singly ionized species.
One might expect only
the UV {\it neutral} gases to be spatially coincident 
with the 21-cm absorbing gas (unless
the 21-cm gas is primordial or has very low heavy element abundances).
However, it was clear in our sample that velocity structure of neutral
UV species was followed closely by singly ionized species as well. This suggests
that ionization fraction and abundance may not change significantly
along a complex. As shown by \cite{2003ApJ...582...49P}, this is a
general observation for DLAs. Further, the strongest component in
singly ionized species corresponded well to the strongest component
in neutral species. These observations justify using singly ionized
species as well. As neutral species are very rare, one obtains the
substantial advantage of a much larger sample.  This allowed
us to investigate systematics for individual absorbers for the first time
(see below). 

\paragraph*{Errors.---}
The redshift errors introduced in digitizing the data are $< 4\times
10^{-6}$, (typically $\sim 10^{-6}$), which is about an order of
magnitude smaller than the observed intrinsic scatter on
\dxx. Further, for result 1, the error quoted is a standard deviation
on the mean, directly reflecting the scatter in \dxxavq.  The error in
result 2 was obtained after introducing errors for \zrad\ and $\langle
z_{\rm UV} \rangle$ (\tr{tab:data}).  Although, formally, this gives a
smaller error, when the errors in \dxxavq\ are increased so that
\csn~$=1$, results 1 and 3 for \dxxavtw\ and its error are very
close.  This shows that {\it the scatter in \dxxavq\ dominates any errors
in individual redshift measurements}. For a single absorption complex
this is a systematic error. However, for many complexes this error has
random sign and magnitude.  This error can be accounted for directly
if treated as a statistical error when calculating the mean value of
$x$ for many absorption complexes, as we do for results 1 and 3.

%Although the two results are consistent, formally result 2 has a
%smaller error.  This is misleading: The observed redshift difference
%$z_{\rm UV}-z_{\rm 21}$ between the strongest components of 21-cm and
%UV may be due to a spatial non-coincidence of hydrogen and heavy
%elements.  For a single absorption complex this is a systematic
%error. However, for many complexes this error has random sign and
%magnitude.  This error can be accounted for directly if treated as a
%statistical error when calculating the mean value of $x$ for many
%absorption complexes, as we do for result 1.

\paragraph*{Discussion of results.---}
In \fr{fig:dxxz} there is
considerable scatter in the values for \dxxavq. For
all spectra all optically observed species tend to group
together in their \dxx\ values on one side of zero. This is because
there is significant offset between the single
\zrad\ value and all \zuv\ values in a
system. This suggests that there {\it is} spatial offset between the
21-cm and UV absorbing gases, but, as expected, this is {\it random}
for different absorbers.
The error of result 1 directly reflects this scatter.
A straightforward statistical calculation misses this effect
(small error in result 2). Result 3 is consistent with
result 1 and the value of $s$ provides an estimate of
the line of sight velocity difference $\Delta v_{\rm los} \sim
cs = 6$~\kmps. 

One possible physical explanation for the observed offset may be a
large angular size for the emitting 21-cm quasar source, as seen by
the absorber.  A 21-cm sightline can then intersect a cold, neutral
hydrogen cloud with little or no heavy elements,
whilst a UV/optical sightline can intersect another cloud with heavy
elements at quite a different velocity.  A large angular size is due to the
combined effects of proximity of the absorber to the quasar (small
\zem~$-$~\zav) and intrinsic size of the radio emitting region.
Note also the good agreement with the 
velocity difference ($\sim 10$~\kmps) 
between 21-cm and mm absorption lines
likely due
to small scale motion of the interstellar medium \cite{2000PhRvL..85.5511C}.

\paragraph*{Comparison with previous results.---}
\cite{1995ApJ...453..596C} used neutral carbon \cone, \cone$^{\ast}$,
lines in the Keck/HIRES absorption spectrum of \otto, to obtain
\dxx~$=(0.70\pm0.55)\times 10^{-5}$. For the same object we obtain
\dxx~$=(-2.59\pm0.74_{\rm stat}\pm1.90_{\rm syst})\times 10^{-5}$.
The central value in \cite{1995ApJ...453..596C} differs from ours
because these authors used a \zuv\ value which is a weighted mean for
the observed components \cite{1994Natur.371...43S}, thus obtaining a
value very close to \zrad\ for this absorption system.  In our
VLT/UVES spectrum of \otto, we detected 23 distinct UV heavy element
transitions whose strongest component was well defined within a few
\kmps.  Further, our use of eight objects allows us to quantify the
systematics due to the sightline issues explained above.  The error in
\cite{1995ApJ...453..596C} is exclusively statistical, as it is based
on a single absorption system. Although at face value this error is
lower than ours, it inevitably contains no information on systematics,
which, as our error estimate shows, dominate.

\paragraph*{Robustness.---}
We stress that a non-zero \dxx\ value is not corroborated by the
sample as a whole, for which the result is robust. For our quoted
result we have not used \zuv\ obtained from \catwo\ whose ionization
potentials are least similar to those for all other elements
used. Even so, the result changes by about $4\%$ if \catwo\ is
included. If we use \zrad\ from the literature, rather than values
determined from our digitized plots, we obtain
\dxxavtw~$=(1.51\pm1.04)\times 10^{-5}$\footnote{For \ztt\ and \oot\
results based on digitized plots have been used in this calculation as
there are no available literature values.}.  If, additionally, we do
not use results for \ztt\ or \oot, thus excluding any results based on
digitized plots, we obtain \dxxavtw~$=(1.77\pm1.12)\times 10^{-5}$. In
all cases $\Delta v_{\rm los}$ remains at 6~\kmps\
\footnote{All results here were obtained using method 3.}.
%(weighted mean
%with increased error bar).}.

\paragraph*{Variation of $\mu$.---}
%???? 
Measurements of $x\equiv \alpha^2 g_p m_e/m_p$ are sensitive to
variation of several fundamental constants. The proton mass $m_p$ is
proportional to the quantum chromodynamic (QCD) scale $\Lambda_{\rm
QCD}$.
%(defined as the position of the Landau pole in the logarithm for
%the running strong coupling constant $\alpha_s(r) \sim
%1/\ln{(\Lambda_{\rm QCD}r)}$).  
The dependence of the proton mass on the
quark mass $m_q$ is very weak ($\Delta m_p/m_p \approx 0.05 \Delta
m_q/m_q$ \cite{2004Flambaum}) and may be neglected. The dependence of
the proton magnetic $g_p$ factor on the fundamental constants is also
quite weak: $\Delta g_p/g_p \approx -0.1 \frac{\Delta
(m_q/\Lambda_{\rm QCD})} {(m_q/\Lambda_{\rm QCD})}$
\cite{2004Flambaum}. Therefore, the most important effects are due to
the variations of $\alpha$ and $\mu\equiv m_e/m_p \approx 0.2
m_e/\Lambda_{\rm QCD}$.  From the definition of $x$ one obtains
$\frac{\mu_z-\mu_0}{\mu_0} \equiv \Delta\mu/\mu =\Delta x/x -
2\Delta\alpha/\alpha -\Delta g_p/g_p$. Using the \daa\ results of
\cite{2003MNRAS.345..609M} and \cite{2004A&A...417..853C}, we obtain 
$\Delta\mu/\mu=(2.31\pm1.03)\times 10^{-5}$ and
$\Delta\mu/\mu=(1.29\pm1.01)\times 10^{-5}$, respectively,
assuming $\Delta g_p/g_p \sim 0$.  
%????
Our result contradicts $\Delta\mu/\mu=(-2.97\pm0.74)\times 10^{-5}$
\cite{2004CR_Petitjean}\footnote{\cite{2004CR_Petitjean} note that some 
systematics are probably hidden, in
particular in the laboratory wavelengths.} but is consistent with
$\Delta \mu/\mu=(0.5\pm3.6)\times 10^{-5}$, ($2\sigma$)
\cite{2004PRL...92...101302U}. Note that our result on $\Delta\mu/\mu$
is derived using a completely independent method compared to
\cite{2004PRL...92...101302U,2004CR_Petitjean}.

A more sophisticated analysis, involving fitting Voigt profiles, would
not significantly affect our results because the uncertainties in our
estimates of the UV redshifts are small compared to the offset between
\zrad\ and \zuv. While more data will improve our estimate of $\Delta
v_{\rm los}$, a sample of $\sim 100$ 21-cm/UV absorbers is required
before averaged line-of-sight velocity differences and individual
redshift measurement errors are similar.

We thank A.~Wolfe for data and useful discussions.

% This
%research has made use of the NASA/IPAC Extragalactic Database (NED)
%which is operated by the Jet Propulsion Laboratory, California
%Institute of Technology, under contract with the National Aeronautics
%and Space Administration. Some of the observations were made with ESO
%telescopes on Paranal. 

%\bibliography{radopt}% Produces the bibliography via BibTeX.

\begin{thebibliography}{31}
\expandafter\ifx\csname natexlab\endcsname\relax\def\natexlab#1{#1}\fi
\expandafter\ifx\csname bibnamefont\endcsname\relax
  \def\bibnamefont#1{#1}\fi
\expandafter\ifx\csname bibfnamefont\endcsname\relax
  \def\bibfnamefont#1{#1}\fi
\expandafter\ifx\csname citenamefont\endcsname\relax
  \def\citenamefont#1{#1}\fi
\expandafter\ifx\csname url\endcsname\relax
  \def\url#1{\texttt{#1}}\fi
\expandafter\ifx\csname urlprefix\endcsname\relax\def\urlprefix{URL }\fi
\providecommand{\bibinfo}[2]{#2}
\providecommand{\eprint}[2][]{\url{#2}}

\bibitem[{\citenamefont{{Uzan}}(2003)}]{2003UzanRMP}
\bibinfo{author}{\bibfnamefont{J.-P.} \bibnamefont{{Uzan}}},
  \bibinfo{journal}{\rmp} \textbf{\bibinfo{volume}{75}}, \bibinfo{pages}{403}
  (\bibinfo{year}{2003}).

\bibitem[{\citenamefont{{Dzuba} et~al.}(1999)\citenamefont{{Dzuba}, {Flambaum},
  and {Webb}}}]{1999PhRvL..82..888D}
\bibinfo{author}{\bibfnamefont{V.~A.} \bibnamefont{{Dzuba}}},
  \bibinfo{author}{\bibfnamefont{V.~V.} \bibnamefont{{Flambaum}}},
  \bibnamefont{and} \bibinfo{author}{\bibfnamefont{J.~K.}
  \bibnamefont{{Webb}}}, \bibinfo{journal}{\prl} \textbf{\bibinfo{volume}{82}},
  \bibinfo{pages}{888} (\bibinfo{year}{1999}).

\bibitem[{\citenamefont{{Webb {\it et al.}}}(1999)}]{1999PhRvL..82..884W}
\bibinfo{author}{\bibfnamefont{J.~K.} \bibnamefont{{Webb {\it et al.}}}},
  \bibinfo{journal}{\prl} \textbf{\bibinfo{volume}{82}}, \bibinfo{pages}{884}
  (\bibinfo{year}{1999}).

\bibitem[{\citenamefont{{Webb {\it et al.}}}(2001)}]{2001PhRvL..87i1301W}
\bibinfo{author}{\bibfnamefont{J.~K.} \bibnamefont{{Webb {\it et al.}}}},
  \bibinfo{journal}{Physical Review Letters} \textbf{\bibinfo{volume}{87}},
  \bibinfo{pages}{091301} (\bibinfo{year}{2001}).

\bibitem[{\citenamefont{{Murphy} et~al.}(2003)\citenamefont{{Murphy}, {Webb},
  and {Flambaum}}}]{2003MNRAS.345..609M}
\bibinfo{author}{\bibfnamefont{M.~T.} \bibnamefont{{Murphy}}},
  \bibinfo{author}{\bibfnamefont{J.~K.} \bibnamefont{{Webb}}},
  \bibnamefont{and} \bibinfo{author}{\bibfnamefont{V.~V.}
  \bibnamefont{{Flambaum}}}, \bibinfo{journal}{\mnras}
  \textbf{\bibinfo{volume}{345}}, \bibinfo{pages}{609} (\bibinfo{year}{2003}).

\bibitem[{\citenamefont{{Chand {\it et al.}}}(2004)}]{2004A&A...417..853C}
\bibinfo{author}{\bibfnamefont{H.}~\bibnamefont{{Chand {\it et al.}}}},
  \bibinfo{journal}{\aap} \textbf{\bibinfo{volume}{417}}, \bibinfo{pages}{853}
  (\bibinfo{year}{2004}).

\bibitem[{\citenamefont{{Srianand {\it et al.}}}(2004)}]{2004PhRvL..92l1302S}
\bibinfo{author}{\bibfnamefont{R.}~\bibnamefont{{Srianand {\it et al.}}}},
  \bibinfo{journal}{\prl} \textbf{\bibinfo{volume}{92}},
  \bibinfo{pages}{121302} (\bibinfo{year}{2004}).

\bibitem[{\citenamefont{{Thompson}}(1975)}]{1975ApL....16....3T}
\bibinfo{author}{\bibfnamefont{R.~I.} \bibnamefont{{Thompson}}},
  \bibinfo{journal}{\aplett} \textbf{\bibinfo{volume}{16}}, \bibinfo{pages}{3}
  (\bibinfo{year}{1975}).

\bibitem[{\citenamefont{{Foltz} et~al.}(1988)\citenamefont{{Foltz}, {Chaffee},
  and {Black}}}]{1988ApJ...324..267F}
\bibinfo{author}{\bibfnamefont{C.~B.} \bibnamefont{{Foltz}}},
  \bibinfo{author}{\bibfnamefont{F.~H.} \bibnamefont{{Chaffee}}},
  \bibnamefont{and} \bibinfo{author}{\bibfnamefont{J.~H.}
  \bibnamefont{{Black}}}, \bibinfo{journal}{\apj}
  \textbf{\bibinfo{volume}{324}}, \bibinfo{pages}{267} (\bibinfo{year}{1988}).

\bibitem[{\citenamefont{{Varshalovich} and
  {Levshakov}}(1993)}]{1993JETPL.....58..237V}
\bibinfo{author}{\bibfnamefont{D.~A.} \bibnamefont{{Varshalovich}}}
  \bibnamefont{and} \bibinfo{author}{\bibfnamefont{S.~A.}
  \bibnamefont{{Levshakov}}}, \bibinfo{journal}{J.~Exp.~Theor.~Phys.~Lett.}
  \textbf{\bibinfo{volume}{58}}, \bibinfo{pages}{237} (\bibinfo{year}{1993}).

\bibitem[{\citenamefont{{Ivanchik {\it et al.}}}(2002)}]{2002AstL...28..423I}
\bibinfo{author}{\bibfnamefont{A.~V.} \bibnamefont{{Ivanchik {\it et al.}}}},
  \bibinfo{journal}{Astronomy Letters} \textbf{\bibinfo{volume}{28}},
  \bibinfo{pages}{423} (\bibinfo{year}{2002}).

\bibitem[{\citenamefont{{Ubachs} and
  {Reinhold}}(2004)}]{2004PRL...92...101302U}
\bibinfo{author}{\bibfnamefont{W.}~\bibnamefont{{Ubachs}}} \bibnamefont{and}
  \bibinfo{author}{\bibfnamefont{E.}~\bibnamefont{{Reinhold}}},
  \bibinfo{journal}{\prl} \textbf{\bibinfo{volume}{92}},
  \bibinfo{pages}{101302} (\bibinfo{year}{2004}).

\bibitem[{\citenamefont{{Curran {\it et al.}}}(2005)}]{2005MNRAS.356.1509C}
\bibinfo{author}{\bibfnamefont{S.}~\bibnamefont{{Curran {\it et al.}}}},
  \bibinfo{journal}{\mnras} \textbf{\bibinfo{volume}{356}},
  \bibinfo{pages}{1509} (\bibinfo{year}{2005}).

\bibitem[{\citenamefont{{Tubbs} and {Wolfe}}(1980)}]{1980ApJ...236L.105T}
\bibinfo{author}{\bibfnamefont{A.~D.} \bibnamefont{{Tubbs}}} \bibnamefont{and}
  \bibinfo{author}{\bibfnamefont{A.~M.} \bibnamefont{{Wolfe}}},
  \bibinfo{journal}{\apj} \textbf{\bibinfo{volume}{236}}, \bibinfo{pages}{L105}
  (\bibinfo{year}{1980}).

\bibitem[{\citenamefont{{Wolfe} and {Davis}}(1979)}]{1979AJ.....84..699W}
\bibinfo{author}{\bibfnamefont{A.~M.} \bibnamefont{{Wolfe}}} \bibnamefont{and}
  \bibinfo{author}{\bibfnamefont{M.~M.} \bibnamefont{{Davis}}},
  \bibinfo{journal}{\aj} \textbf{\bibinfo{volume}{84}}, \bibinfo{pages}{699}
  (\bibinfo{year}{1979}).

\bibitem[{\citenamefont{{Cowie} and {Songaila}}(1995)}]{1995ApJ...453..596C}
\bibinfo{author}{\bibfnamefont{L.~L.} \bibnamefont{{Cowie}}} \bibnamefont{and}
  \bibinfo{author}{\bibfnamefont{A.}~\bibnamefont{{Songaila}}},
  \bibinfo{journal}{\apj} \textbf{\bibinfo{volume}{453}}, \bibinfo{pages}{596}
  (\bibinfo{year}{1995}).

\bibitem[{\citenamefont{{Wolfe} et~al.}(1981)\citenamefont{{Wolfe}, {Briggs},
  and {Jauncey}}}]{1981ApJ...248..460W}
\bibinfo{author}{\bibfnamefont{A.~M.} \bibnamefont{{Wolfe}}},
  \bibinfo{author}{\bibfnamefont{F.~H.} \bibnamefont{{Briggs}}},
  \bibnamefont{and} \bibinfo{author}{\bibfnamefont{D.~L.}
  \bibnamefont{{Jauncey}}}, \bibinfo{journal}{\apj}
  \textbf{\bibinfo{volume}{248}}, \bibinfo{pages}{460} (\bibinfo{year}{1981}).

\bibitem[{\citenamefont{{Tzanavaris {\it et al.}}}(2004)}]{2004Tzan}
\bibinfo{author}{\bibfnamefont{P.}~\bibnamefont{{Tzanavaris {\it et al.}}}}
  (\bibinfo{year}{2004}), \eprint{astro-ph/0412649, online preprint version of
  this paper}.

\bibitem[{\citenamefont{{Tzanavaris {\it et al.}}}(2005)}]{2005Tzan}
\bibinfo{author}{\bibfnamefont{P.}~\bibnamefont{{Tzanavaris {\it et al.}}}}
  (\bibinfo{year}{2005}), \eprint{\mnras, in preparation}.

\bibitem[{\citenamefont{{Wolfe {\it et al.}}}(1985)}]{1985ApJ...294L..67W}
\bibinfo{author}{\bibfnamefont{A.~M.} \bibnamefont{{Wolfe {\it et al.}}}},
  \bibinfo{journal}{\apj} \textbf{\bibinfo{volume}{294}}, \bibinfo{pages}{L67}
  (\bibinfo{year}{1985}).

\bibitem[{\citenamefont{{Kanekar} and
  {Chengalur}}(2001{\natexlab{a}})}]{2001AA...369...42K}
\bibinfo{author}{\bibfnamefont{N.}~\bibnamefont{{Kanekar}}} \bibnamefont{and}
  \bibinfo{author}{\bibfnamefont{J.~N.} \bibnamefont{{Chengalur}}},
  \bibinfo{journal}{\aap} \textbf{\bibinfo{volume}{369}}, \bibinfo{pages}{42}
  (\bibinfo{year}{2001}{\natexlab{a}}).

\bibitem[{\citenamefont{{Kanekar} and
  {Chengalur}}(2001{\natexlab{b}})}]{2001MNRAS.325..631K}
\bibinfo{author}{\bibfnamefont{N.}~\bibnamefont{{Kanekar}}} \bibnamefont{and}
  \bibinfo{author}{\bibfnamefont{J.~N.} \bibnamefont{{Chengalur}}},
  \bibinfo{journal}{\mnras} \textbf{\bibinfo{volume}{325}},
  \bibinfo{pages}{631} (\bibinfo{year}{2001}{\natexlab{b}}).

\bibitem[{\citenamefont{{Brown} and {Spencer}}(1979)}]{1979ApJ...230L...1B}
\bibinfo{author}{\bibfnamefont{R.~L.} \bibnamefont{{Brown}}} \bibnamefont{and}
  \bibinfo{author}{\bibfnamefont{R.~E.} \bibnamefont{{Spencer}}},
  \bibinfo{journal}{\apj} \textbf{\bibinfo{volume}{230}}, \bibinfo{pages}{L1}
  (\bibinfo{year}{1979}).

\bibitem[{\citenamefont{{Wolfe {\it et al.}}}(1978)}]{1978ApJ...222..752W}
\bibinfo{author}{\bibfnamefont{A.~M.} \bibnamefont{{Wolfe {\it et al.}}}},
  \bibinfo{journal}{\apj} \textbf{\bibinfo{volume}{222}}, \bibinfo{pages}{752}
  (\bibinfo{year}{1978}).

\bibitem[{\citenamefont{{Lanzetta} and {Bowen}}(1992)}]{1992ApJ...391...48L}
\bibinfo{author}{\bibfnamefont{K.~M.} \bibnamefont{{Lanzetta}}}
  \bibnamefont{and} \bibinfo{author}{\bibfnamefont{D.~V.}
  \bibnamefont{{Bowen}}}, \bibinfo{journal}{\apj}
  \textbf{\bibinfo{volume}{391}}, \bibinfo{pages}{48} (\bibinfo{year}{1992}).

\bibitem[{\citenamefont{{Wolfe}}()}]{wolfep}
\bibinfo{author}{\bibfnamefont{A.~M.} \bibnamefont{{Wolfe}}}, \eprint{private
  communication}.

\bibitem[{\citenamefont{{Prochaska}}(2003)}]{2003ApJ...582...49P}
\bibinfo{author}{\bibfnamefont{J.~X.} \bibnamefont{{Prochaska}}},
  \bibinfo{journal}{\apj} \textbf{\bibinfo{volume}{582}}, \bibinfo{pages}{49}
  (\bibinfo{year}{2003}).

\bibitem[{\citenamefont{{Carilli {\it et al.}}}(2000)}]{2000PhRvL..85.5511C}
\bibinfo{author}{\bibfnamefont{C.~L.} \bibnamefont{{Carilli {\it et al.}}}},
  \bibinfo{journal}{Physical Review Letters} \textbf{\bibinfo{volume}{85}},
  \bibinfo{pages}{5511} (\bibinfo{year}{2000}).

\bibitem[{\citenamefont{{Songaila {\it et al.}}}(1994)}]{1994Natur.371...43S}
\bibinfo{author}{\bibfnamefont{A.}~\bibnamefont{{Songaila {\it et al.}}}},
  \bibinfo{journal}{\nat} \textbf{\bibinfo{volume}{371}}, \bibinfo{pages}{43}
  (\bibinfo{year}{1994}).

\bibitem[{\citenamefont{{Flambaum {\it et al.}}}(2004)}]{2004Flambaum}
\bibinfo{author}{\bibfnamefont{V.~V.} \bibnamefont{{Flambaum {\it et al.}}}},
  \bibinfo{journal}{\prd} \textbf{\bibinfo{volume}{69}},
  \bibinfo{pages}{115006} (\bibinfo{year}{2004}).

\bibitem[{\citenamefont{{Petitjean {\it et al.}}}(2004)}]{2004CR_Petitjean}
\bibinfo{author}{\bibfnamefont{P.}~\bibnamefont{{Petitjean {\it et al.}}}},
  \bibinfo{journal}{Comptes Rendus Acad.~Sci. (Paris)}
  \textbf{\bibinfo{volume}{5}}, \bibinfo{pages}{411} (\bibinfo{year}{2004}).

\end{thebibliography}

\begin{figure*}[htb]
\includegraphics[width=1.5\columnwidth, angle=0]{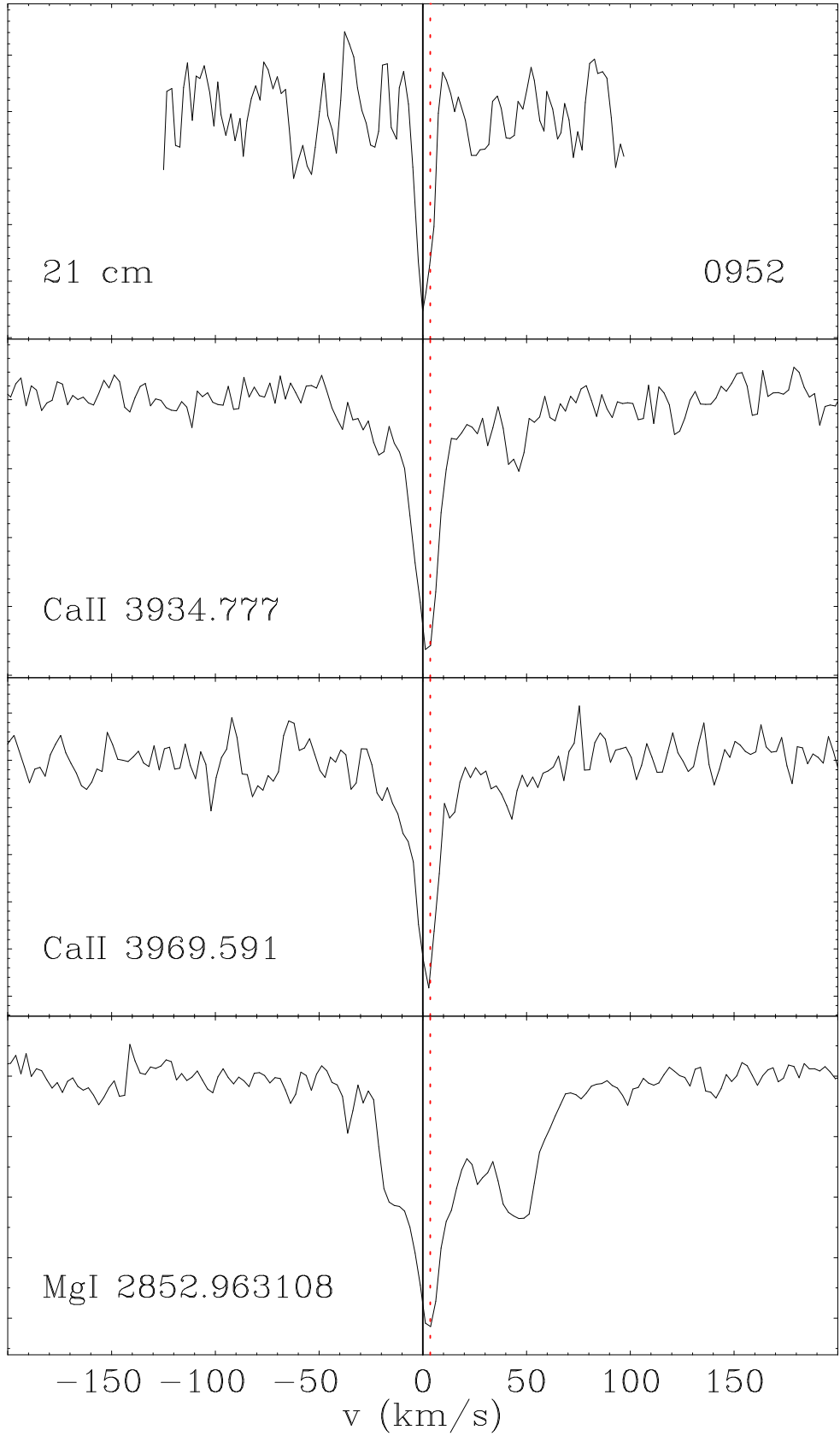}
\caption{\label{fig:0952} Velocity plot for 21-cm and UV absorption
towards quasar \znf.  The solid vertical line at 0~\kmps\ is at
\zrad. The dotted vertical line is at $\langle z_{\rm UV}\rangle$.
In this and subsequent plots \catwo\ is shown for illustration only but
has not been used in the calculation of the plotted $\langle z_{\rm UV}\rangle$.}
\end{figure*}

\begin{figure*}[htb]
\includegraphics[width=1.5\columnwidth, angle=0]{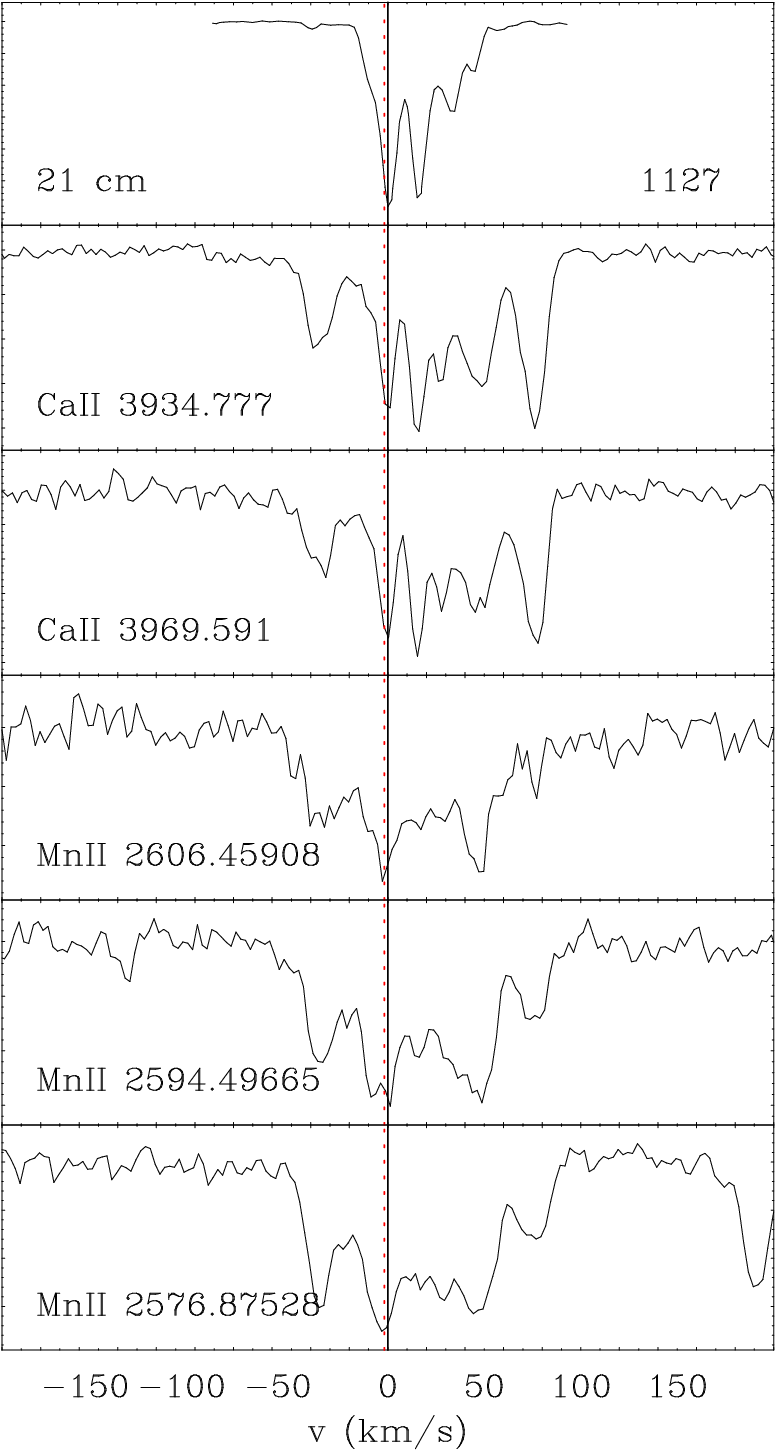}
\caption{\label{fig:1127L} Velocity plot for 21-cm and UV absorption
towards quasar \oot.  The solid vertical line at 0~\kmps\ is at
\zrad. The dotted vertical line is at $\langle z_{\rm UV}\rangle$.}
\end{figure*}

\begin{figure*}[htb]
\includegraphics[width=1.5\columnwidth, angle=0]{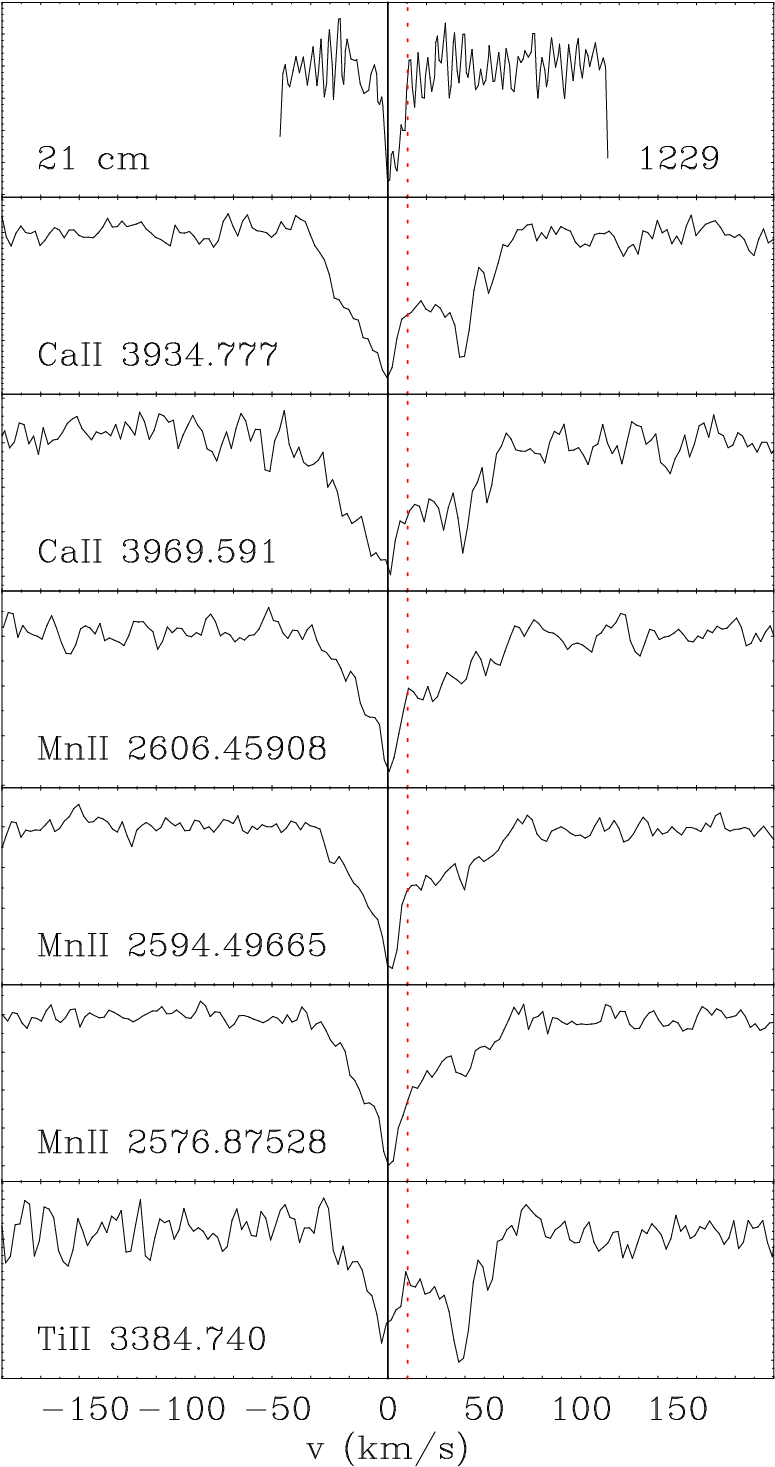}
\caption{\label{fig:1229} Velocity plot for 21-cm and UV absorption
towards quasar \ott.  The solid vertical line at 0~\kmps\ is at
\zrad. The dotted vertical line is at $\langle z_{\rm UV}\rangle$.}
\end{figure*}

\begin{figure*}[htb]
\includegraphics[width=1.5\columnwidth, angle=0]{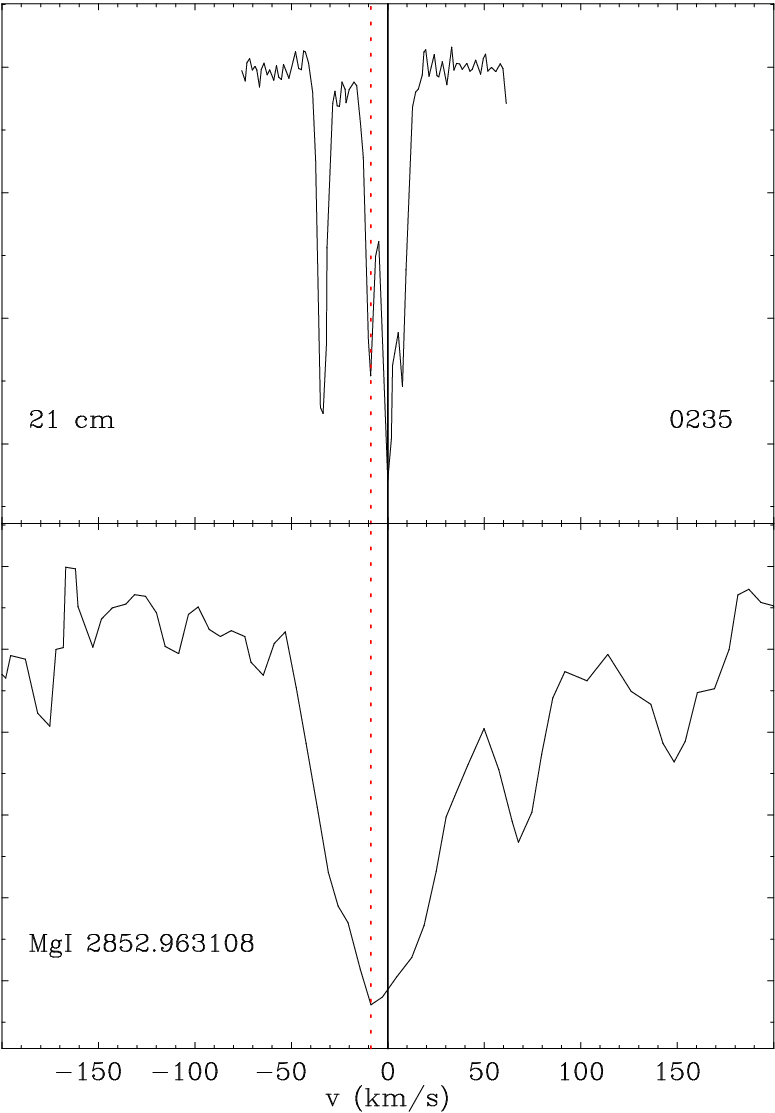}
\caption{\label{fig:0235} Velocity plot for 21-cm and UV absorption
towards quasar \ztt.  The solid vertical line at 0~\kmps\ is at
\zrad. The dotted vertical line is at $\langle z_{\rm UV}\rangle$.}
\end{figure*}

\begin{figure*}[htb]
\includegraphics[width=1.5\columnwidth, angle=0]{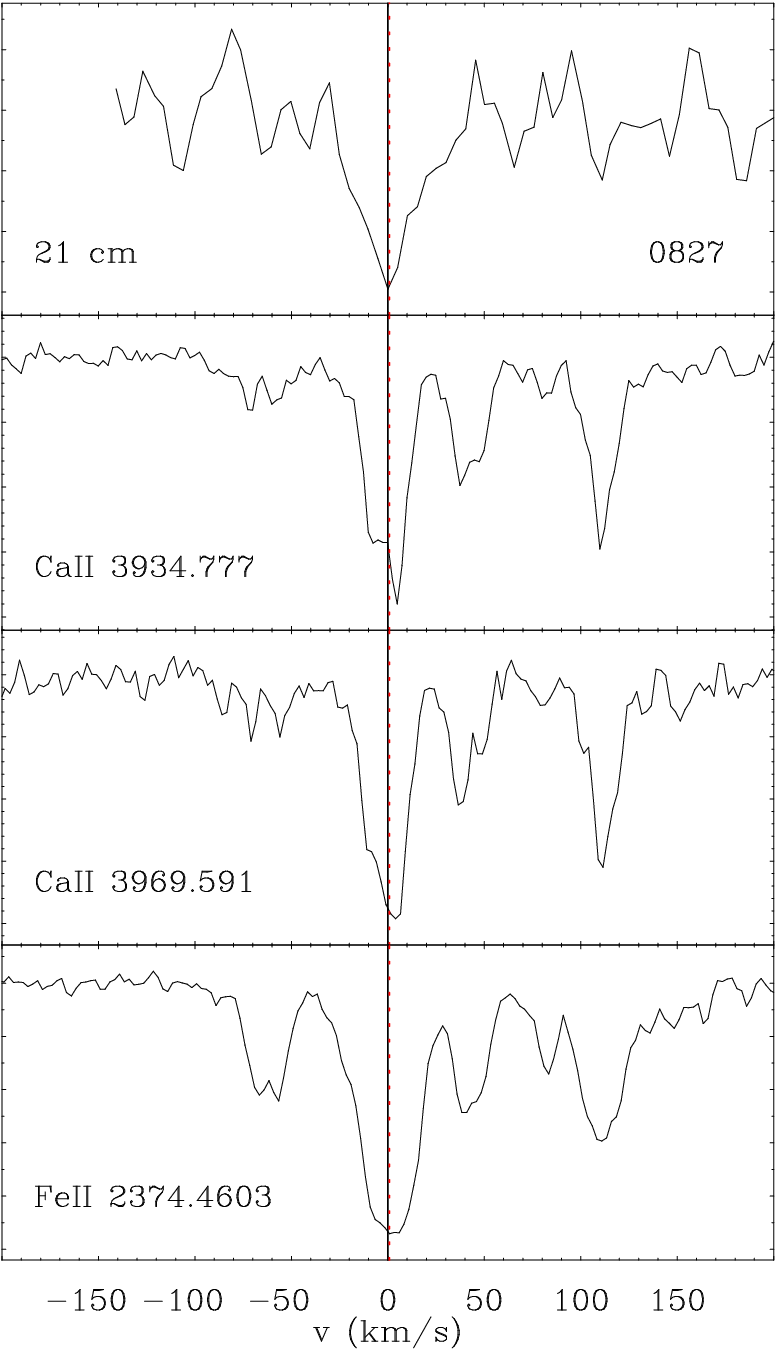}
\caption{\label{fig:0827} Velocity plot for 21-cm and UV absorption
towards quasar \zet.  The solid vertical line at 0~\kmps\ is at
\zrad. The dotted vertical line is at $\langle z_{\rm UV}\rangle$.}
\end{figure*}

\vspace{-2cm}
\begin{figure*}[htb]
\includegraphics[width=1.3\columnwidth, angle=0]{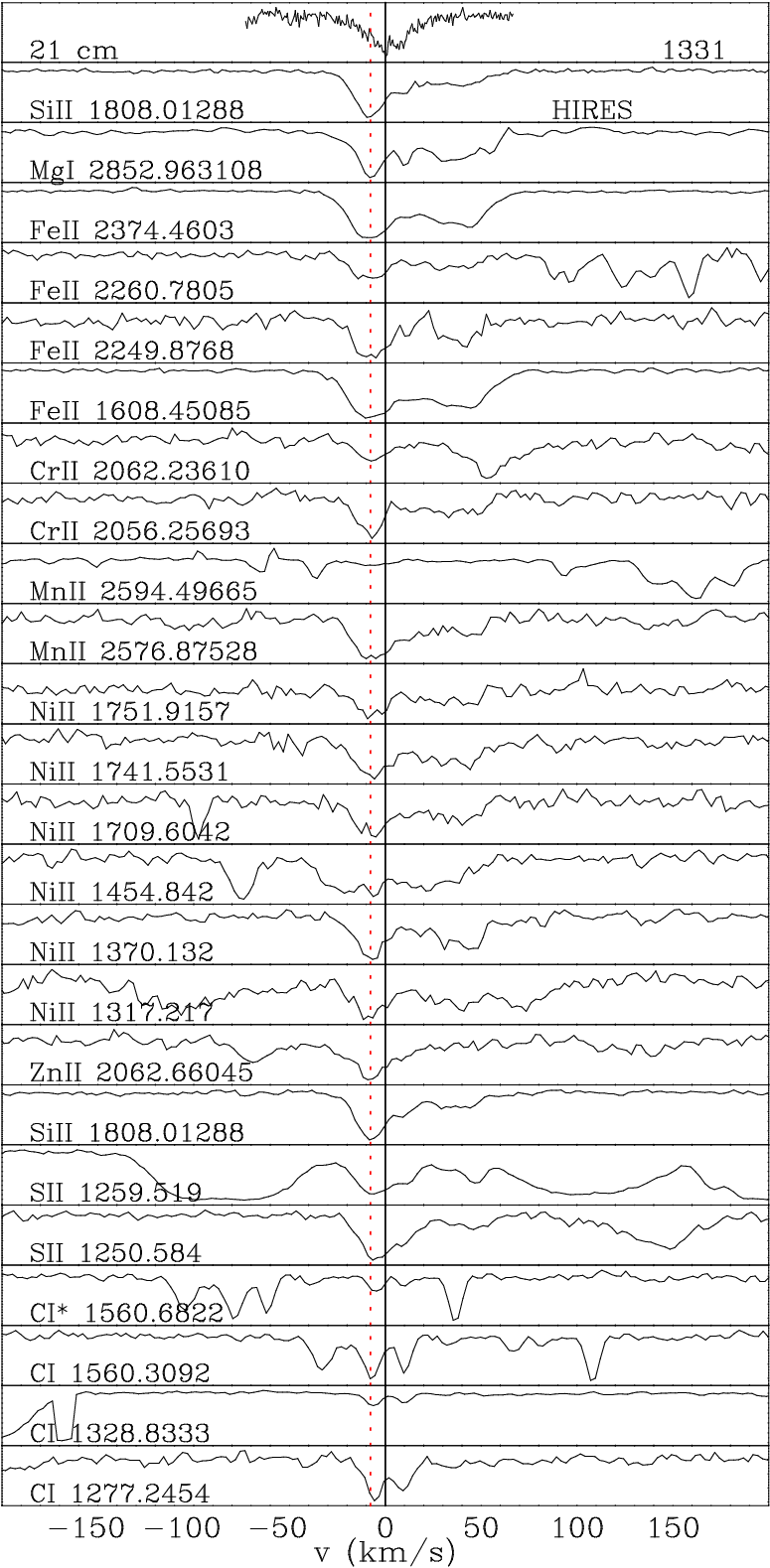}
\caption{\label{fig:1331} Velocity plot for 21-cm and UV absorption
towards quasar \otto.  The solid vertical line at 0~\kmps\ is at
\zrad. The dotted vertical line is at $\langle z_{\rm UV}\rangle$.}
\end{figure*}

\vspace{-2cm}
\begin{figure*}[htb]
\includegraphics[width=1.3\columnwidth, angle=0]{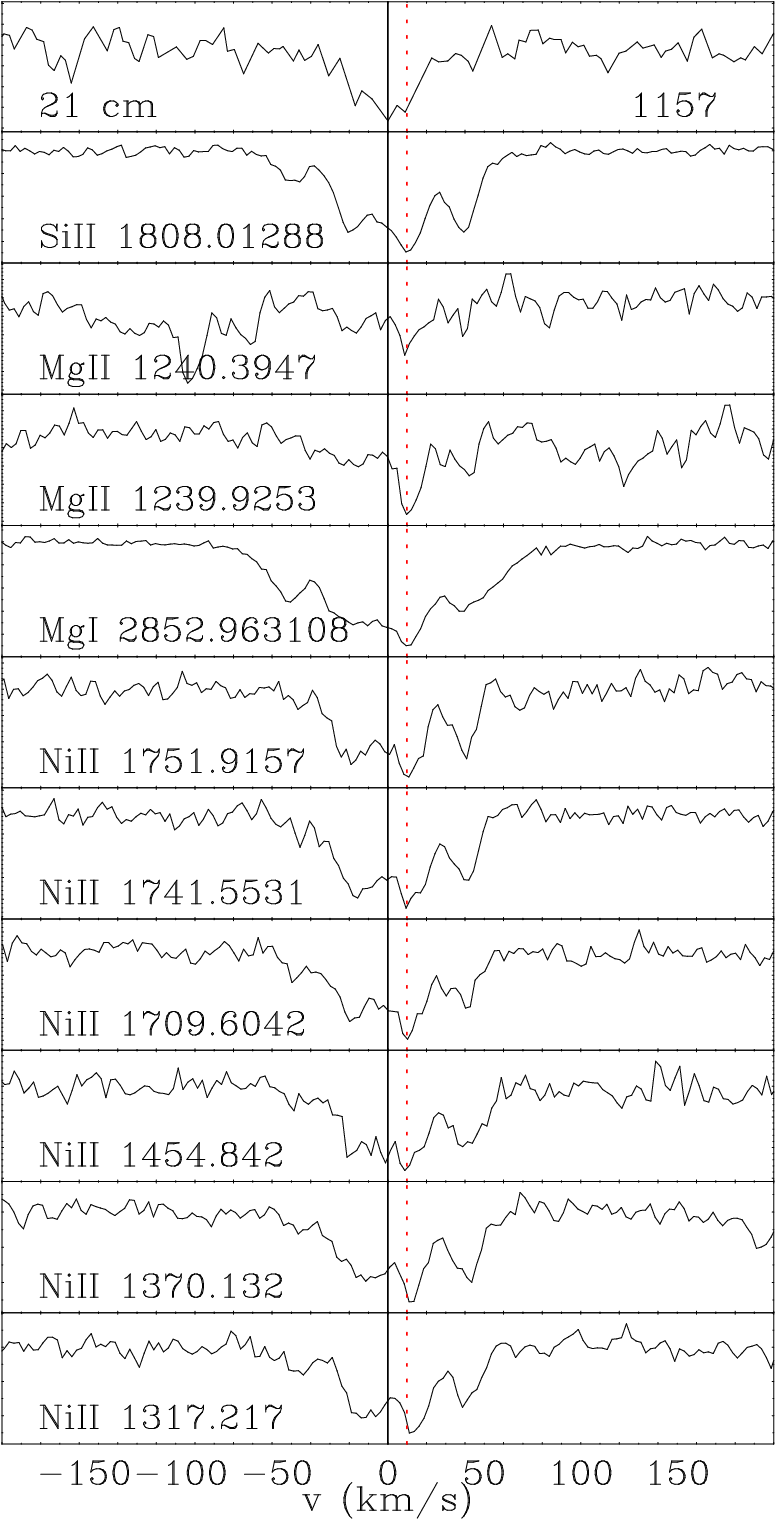}
\caption{\label{fig:1157} Velocity plot for 21-cm and UV absorption
towards quasar \oof.  The solid vertical line at 0~\kmps\ is at
\zrad. The dotted vertical line is at $\langle z_{\rm UV}\rangle$.}
\end{figure*}

\vspace{-2cm}
\begin{figure*}[htb]
\includegraphics[width=1.3\columnwidth, angle=0]{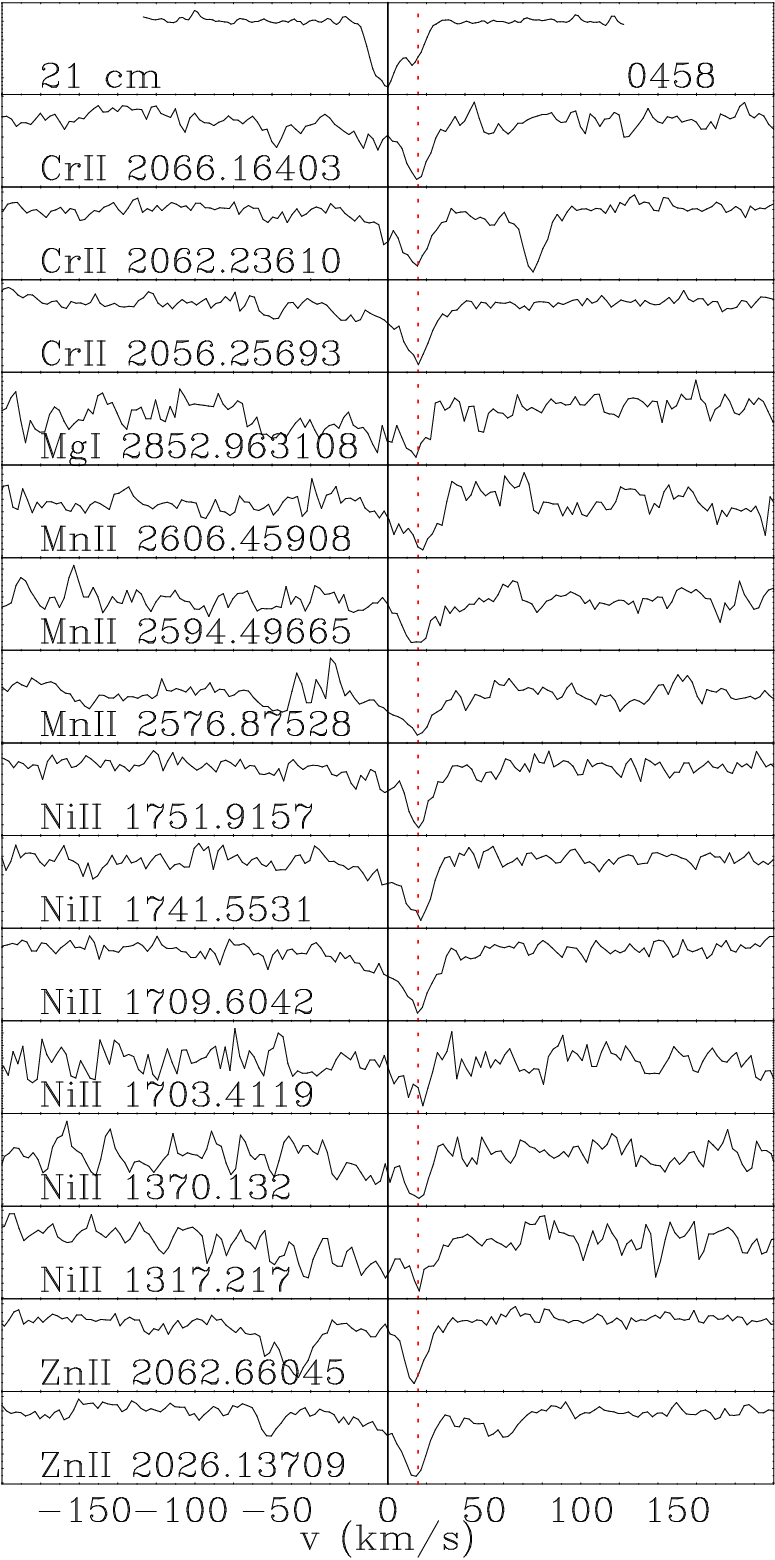}
\caption{\label{fig:0458} Velocity plot for 21-cm and UV absorption
towards quasar \zff.  The solid vertical line at 0~\kmps\ is at
\zrad. The dotted vertical line is at $\langle z_{\rm UV}\rangle$.}
\end{figure*}

\end{document}